\documentclass[preprint2]{aastex6}

\usepackage{natbib}
\begin{document}


\title{Rocky Planetesimal Formation via Fluffy Aggregates of Nanograins
}


\author{Sota Arakawa \altaffilmark{1} and Taishi Nakamoto}
\affil{Department of Earth and Planetary Sciences, Tokyo Institute of Technology, Meguro, Tokyo, 152-8551, Japan}



\altaffiltext{1}{e-mail: arakawa.s.ac@m.titech.ac.jp}

\begin{abstract}
Several pieces of evidence suggest that silicate grains in primitive meteorites are not interstellar grains but condensates formed in the early solar system.
Moreover, the size distribution of matrix grains in chondrites implies that these condensates might be formed as nanometer-sized grains.
Therefore, we propose a novel scenario for rocky planetesimal formation in which nanometer-sized silicate grains are produced by evaporation and recondensation events in early solar nebula, and rocky planetesimals are formed via aggregation of these nanograins.
We reveal that silicate nanograins can grow into rocky planetesimals via direct aggregation without catastrophic fragmentation and serious radial drift, and our results provide a suitable condition for protoplanet formation in our solar system.

\end{abstract}




\section{Introduction}
The standard scenario for planet formation is based on the planetesimal hypothesis, however, the process that caused submicron-sized interstellar dust grains to evolve into kilometer-sized planetesimals is not yet understood, especially for rocky planetesimals.
This is because there are several ``barriers'' for planetesimal formation, e.g., the fragmentation barrier \citep[e.g.,][]{Blum+1993} and the radial drift barrier \citep[e.g.,][]{Weidenschilling1977}.

To avoid these barriers, several mechanisms have been proposed to explain how dust aggregates grow into planetesimals, e.g., gravitational instability of a dust layer \citep[e.g.,][]{Goldreich+1973}, and self-gravitational collapse caused by the two-fluid streaming instability of solids and gas \citep[e.g.,][]{Youdin+2005,Taki+2016}.
Nevertheless, the pathway of rocky planetesimal formation is still unknown.
This is because these mechanisms require meter-sized (or much larger) dust aggregates, and it is difficult for silicate dust to form such large aggregates.
In addition, even if meter-sized dense dust aggregates are formed from submicron-sized interstellar dust grains, these aggregates would likely suffer serious radial drift.

\citet{Okuzumi+2012} and \citet{Kataoka+2013b} revealed that icy planetesimals can be formed via direct aggregation of fluffy icy aggregates in contrast to rocky planetesimals.
If silicate dust aggregates can also grow with high porosity, and if these aggregates can survive high-speed collisions in a turbulent disk, then rocky planetesimals may also be formed via direct aggregation.

If the building blocks of rocky planetesimals are submicron-sized interstellar dust grains, and rocky planetesimals are formed from submicron-sized silicate grains directly, then there are many severe problems for planetesimal formation, including the fragmentation barrier and the radial drift barrier.
However, these barriers can be broken if the building blocks are nanometer-sized grains.
This is because the critical collision velocity for catastrophic disruption depends on the size of monomers \citep{Dominik+1997}, and the critical velocity increases when dust aggregates are constituted by nanograins.
The porosity of dust aggregates may also increase when the monomers of these aggregates are nanometer-sized grains.

Several lines of evidence suggest that silicate grains in meteorites are not the same grains found in the interstellar medium; instead, they were formed via the vapor phase.
From the point of view of isotopic composition, almost all dust grains in our solar system are isotopically homogeneous for most elements, including silicon \citep[e.g.,][]{Chakrabarti+2010b} and magnesium \citep[e.g.,][]{Chakrabarti+2010a}, but not for some volatile elements such as nitrogen \citep[e.g.,][]{Furi+2015}.
These facts suggest that almost all silicate grains in our solar system have experienced evaporation at least once, and that they were formed via condensation from the vapor phase in the early solar nebula.
In addition, infrared spectra of cometary comae in our solar system and circumstellar disks around other young stars also suggest that most of silicate and organic dust grains in protoplanetary disks are not altered interstellar dust grains but the reincarnated ones \citep{Kimura2013}.
Then it is likely that the size of building blocks of rocky planetesimals is different from the size of interstellar dust grains.

We could obtain the information of monomer size by observations of primitive meteorites and interplanetary dust particles (IDPs).
\citet{Toriumi1989} observed the size distribution of matrix grains in the Allende CV3.2 chondrite, and it was revealed that the peak of the size-frequency distribution is located at $5\ \rm{nm}$ in diameter.
Glass with embedded metal and sulfides (GEMS) grains in IDPs, which are regarded as non-equilibrium condensates formed in the early solar nebula, dominantly contain $5$--$50\ \rm{nm}$-sized kamacite and Fe-Ni sulfide grains in an amorphous silicate matrix \citep[e.g.,][]{Keller+2011}.
Most GEMS grains also show highly heterogeneous elemental distributions in chemical maps at similar scale, then, these GEMS grains might be aggregates of nanometer-sized distinct subgrains \citep{Keller+2011}.

Although the size distribution of matrix grains in Allende meteorite seems to have a power-law tail ranging to 10 $\mu$m, this broadening of the size distribution might not represent the nature of monomers but reflect the impact of asteroidal metamorphism.
\citet{Toriumi1989} reported that micron-sized grains show the sintered microstructure but nanometer-sized grains display little evidence of sintering.
\citet{Ashworth1977} measured the size-frequency distribution of ordinary chondrites whose metamorphic grades are different, and revealed that the size of matrix grains clearly shows the effect of grain growth by asteroidal metamorphism.
Furthermore, the theory of Brownian coagulation predicts that the initial size distribution of condensates must be very approximately log-normal \citep[e.g.,][]{Friedlander+1966}.
Then, the initial size frequency of matrix grains might have a peak at several nanometers not only for the number but also for the mass and for the binding energy.

Therefore, we must test a scenario in which the building blocks of rocky planetesimals are not submicron-sized interstellar dust grains but nanometer-sized grains formed via evaporation and condensation of silicate dust.

We notice a bow shock produced by an icy planetesimal revolving in a highly eccentric orbit in the early solar nebula \citep[e.g.,][]{Nagasawa+2014} as a possible site of nanograin formation.
Planetesimal bow shocks can evaporate submicron- and even micron-sized silicate grains completely behind shock fronts \citep{Miura+2005}.
In addition, planetesimal bow shocks can produce nanometer-sized grains via recondensation of silicate grains \citep{Miura+2010}.
The radius of grains formed by condensation of evaporated silicate dust is roughly proportional to the density of evaporated dust and the cooling timescale of evaporated dust \citep{Yamamoto+1977}.
Therefore, not only planetesimal bow shocks but also other short-timescale heating events in the solar nebula \citep[e.g., lightning discharge in protoplanetary disk;][]{Muranushi2010} have the potential to form nanograins.

In this study, we assume an initial condition in which all the dust grains in the inner region of the solar nebula are nanometer-sized monomers, for simplicity.
Of course, in reality, at least some grains have not experienced evaporation and recondensation in the solar nebula, and they may be called presolar grains; however, the abundance of presolar grains is very low.
This situation can be realized when all the silicate grains in the inner region of the solar nebula are evaporated by heating events such as planetesimal bow shocks.
We confirm that not only icy aggregates but also rocky aggregates break the fragmentation barrier if the rocky aggregates are composed of nanometer-sized grains.
We also calculate how much the dust aggregates constructed from nanograins are compressed in the solar nebula, and investigate whether the growth is rapid enough to overcome the radial drift barrier by comparing the timescales of growth and radial drift.

\section{Model}
We assume the minimum mass solar nebula \citep{hayashi1981}.
The gas surface density $\Sigma_{\rm{g}}$ and the dust surface density $\Sigma_{\rm{d}}$ are $\Sigma_{\rm{g}} = 1700 {(R / 1\ \rm{au})}^{- 3/2}\ \rm{g}\ \rm{cm}^{-2}$ and $\Sigma_{\rm{d}} = 7.1 {(R / 1\ \rm{au})}^{- 3/2}\ \rm{g}\ \rm{cm}^{-2}$, respectively, where $R$ is the distance from the Sun.
The temperature profile is $280 {(R / 1\ \rm{au})}^{- 1/2}\ \rm{K}$, and the $\rm{H}_{2}\rm{O}$ snowline is located at $2.7\ \rm{au}$ in the adopted disk model.

The motion of dust aggregates in the gas disk is induced by Brownian motion, radial drift, azimuthal motion, and turbulence.
Here we use the analytic formula for Kolmogorov turbulence \citep{Ormel+2007}.
The timescale of the largest eddies $t_{\rm{L}}$ is $t_{\rm{L}} = {\Omega_{\rm{K}}}^{-1}$, where $\Omega_{\rm{K}}$ is the orbital angular velocity, and the timescale of the smallest eddies $t_{\eta}$ is $t_{\eta} = {\rm{Re}_{\rm{t}}}^{- 1/2} t_{\rm{L}}$, where $\rm{Re}_{\rm{t}}$ is the turbulent Reynolds number \citep{Cuzzi+2003}.

We assume that the radius of nanograins $a_{0}$ is $a_{0} = 2.5\ \rm{nm}$; then, the rolling energy $E_{\rm{roll}} = 6 {\pi}^{2} \gamma a_{0} \xi = 1.1 \times 10^{-11}\ \rm{erg}$, where $\gamma = 25\ \rm{erg}\ \rm{cm}^{-2}$ is the surface energy of silicate dust, and $\xi = 0.3\ \rm{nm}$ is the critical displacement \citep{Wada+2007}.
The critical displacement $\xi$ has an uncertainty from $0.2\ \rm{nm}$ to $3.2\ \rm{nm}$ \citep{heim+1999}, and we assume $\xi = 0.3\ \rm{nm}$ in this study.

For simplicity, we do not consider the mass and density distribution of aggregates in this study.
The validity of using this approximation is ensured by numerical simulations that considered the size distribution of dust aggregates \citep[e.g.,][]{Okuzumi+2012}.
Additionally, we assumed that even if the stopping time of aggregates $t_{\rm{s}}$ is shorter than the timescale of the smallest eddies $t_{\eta}$, the collision velocity induced by turbulence is not zero, as expected based on the formula given by \citet{Ormel+2007}, but one-tenth the turbulence-induced relative velocity between the dust aggregate and gas, considering that dust aggregates have mass and density distribution in reality \citep[e.g.,][]{Okuzumi+2011}.

We calculate the denisity evolution of dust aggregates by considering (i) the hit-and-stick of two colliding aggregates without compression, (ii) collisional compression caused by high-speed collisions, (iii) static compression caused by the ram pressure of the disk gas, and (iv) static compression caused by self-gravity.

The structure of dust aggregates formed by hit-and-stick growth is highly porous, and the fractal dimension of these aggregates is approximately two.
The density of aggregates formed by hit-and-stick growth $\rho_{\rm{hit}}$ is given by
\begin{equation}
\rho_{\rm{hit}} = {\left( \frac{3}{5} \right)}^{3/2} N^{- 1/2} \rho_{0},
\end{equation}
where $N \ge 2$ is the number of monomer nanograins in an aggregate, and $\rho_{0} = 3\ \rm{g}\ \rm{cm}^{-3}$ is the material density of silicate dust \citep[e.g.,][]{Wada+2008}.
The density decreases with the mass of the power of $- 1/2$ when dust aggregates grow by hit-and-stick.

When the impact energy of two colliding aggregates $E_{\rm{imp}}$ reaches the rolling energy $E_{\rm{roll}}$, the dust aggregate formed by collisions has a higher density than that of an aggregate formed by hit-and-stick growth.
The density of aggregates compressed by collision $\rho_{\rm{col}}$ is given by
\begin{equation}
\rho_{\rm{col}} = {\left( \frac{E_{\rm{imp}}}{b E_{\rm{roll}}} \right)}^{3/10} \rho_{\rm{hit}},
\end{equation}
where $b = 0.15$ is a dimensionless constant obtained by numerical simulations \citep{Wada+2008}.
The impact energy of two equal-mass aggregates $E_{\rm{imp}}$ is $E_{\rm{imp}} = {(1/8)} m {\Delta v}^{2}$, where $m$ is the total mass of the two aggregates, and ${\Delta v}$ is the collision velocity.
We assume that ${\Delta v}$ is the root sum square of Brownian motion and turbulence motion.

The equilibrium density of a dust aggregate formed by static compression is given by \citet{Kataoka+2013a}.
We consider the sources of the static compression to be the ram pressure of the disk gas and the self-gravity of the large aggregate.
The equilibrium density of gas compression $\rho_{\rm{gas}}$ is given by
\begin{equation}
\rho_{\rm{gas}} = {\left( \frac{{a_{0}}^{3}}{E_{\rm{roll}}} \frac{m v}{\pi r^{2} t_{\rm{s}}} \right)}^{1/3} \rho_{0},
\end{equation}
where $r$ is the radius of the aggregate, and $t_{\rm{s}}$ is the stopping time \citep{Kataoka+2013b}.
We use the gas drag law derived by \citet{Weidenschilling1977} to obtain the stopping time $t_{\rm{s}}$.
The rolling energy $E_{\rm{roll}}$ is proportional to the critical displacement $\xi$, and the uncertainty of the dust density derived from the uncertainty of $\xi$ is minor because the equilibrium dust density is proportional to only ${E_{\rm{roll}}}^{1/3}$.

Similarly, the equilibrium density of self-gravitational compression $\rho_{\rm{grav}}$ is given by
\begin{equation}
\rho_{\rm{grav}} = {\left( \frac{{a_{0}}^{3}}{E_{\rm{roll}}} \frac{G m^{2}}{\pi r^{4}} \right)}^{1/3} \rho_{0},
\end{equation}
where $G$ is the gravitational constant \citep{Kataoka+2013b}.
The equilibrium density of self-gravity compression is independent of the disk properties, and once the gravitational compression is effective, the density increases with the mass of the power of $2/5$.

\section{Results}
In this study, we calculate the pathways of dust aggregate growth in mass-density space.
In addition, we investigate whether the growth of aggregates is rapid enough to avoid the radial drift barrier by comparing the timescales of growth $t_{\rm{grow}}$ and radial drift $t_{\rm{drift}}$.
The timescale of growth $t_{\rm{grow}}$ is defined as $t_{\rm{grow}} = m / {\left( \pi r^{2} \rho_{\rm{d}} {\Delta v} \right)}$, where $\rho_{\rm{d}} = \Sigma_{\rm{d}} / {\left( \sqrt{2 \pi} h_{\rm{d}} \right)}$ is the spacial mass density at the midplane, and $h_{\rm{d}}$ is the dust scaleheight.
The dust scaleheight $h_{\rm{d}}$ is given by \citet{Youdin+2007}.
The timescale of radial drift $t_{\rm{drift}}$ is defined as the orbital radius divided by the radial drift velocity.
We expect that dust aggregates can grow without significant radial drift if the condition $t_{\rm{grow}} < {(1/30)} t_{\rm{drift}}$ is satisfied, which is obtained by numerical simulations \citep{Okuzumi+2012}.

At first, we confirm whether the maximum collision velocity ${\Delta v}_{\rm{cr}}$ satisfies the condition for growth without serious fragmentation.
The maximum collision velocity ${\Delta v}_{\rm{max}}$ is ${\Delta v}_{\rm{max}} \simeq \sqrt{\alpha} c_{\rm{s}}$, where $c_{\rm{s}} = 9.9 \times 10^{4} {(r / 1\ \rm{au})}^{- 1/4}\ \rm{cm}\ \rm{s}^{-1}$ is the sound velocity \citep{Ormel+2007}.
The critical velocity for catastrophic disruption ${\Delta v}_{\rm{cr}}$ is ${\Delta v}_{\rm{cr}} \simeq 6 \times 10^{2}\ \rm{cm}\ \rm{s}^{-1}$ when the radius of the monomer is $100\ \rm{nm}$ \citep{Wada+2009}.
The monomer size dependence of the critical velocity is ${\Delta v}_{\rm{cr}} \propto {a_{0}}^{- 5/6}$ \citep{Dominik+1997}.
Although there are no experiments on collision and growth of dust aggregates constructed from nanometer-sized monomers, we extrapolate this relation for evaluating the critical velocity in this study.
By using these values, we calculate the critical monomer radius for catastrophic disruption as $14\ \rm{nm}$ for the case of $\alpha < 10^{-3}$ and $r = 1\ \rm{au}$, by considering the given requirement, ${\Delta v}_{\rm{cr}} \ge {\Delta v}_{\rm{max}}$.
Therefore, by considering the aggregation of nanograins, dust aggregates can evolve without catastrophic disruption in our calculation.

We now discuss whether dust aggregates can overcome the radial drift barrier.
Figure 1 shows that the pathways of the dust aggregates at $1\ \rm{au}$ and $2.2\ \rm{au}$ overcome the radial drift problem.
We select the orbital radii $R = 1\ \rm{au}$ and $R = 2.2\ \rm{au}$ because the distance from the Sun to Earth is $1\ \rm{au}$ and the distance from the Flora family asteroids, which are considered as convincing candidates for being the parent bodies of L-type ordinary chondrites \citep{Nesvorny+2002}, is approximately $2.2\ \rm{au}$, respectively.
We assume the {\it alpha parameter} $\alpha$ associated with the strength of turbulence \citep{Ormel+2007} is $\alpha = 10^{-4}$ for Figures 1a and 1b, and $\alpha = 10^{-3}$ for Figures 1c and 1d.
Here, we use an approximation stating that the dust aggregates do not have mass distribution.
The growth pathways at $1\ \rm{au}$ are plotted in Figures 1a and 1c, and the pathways at $2.2\ \rm{au}$ are plotted in Figures 1b and 1d.

\begin{figure*}[htb!]
\plottwo{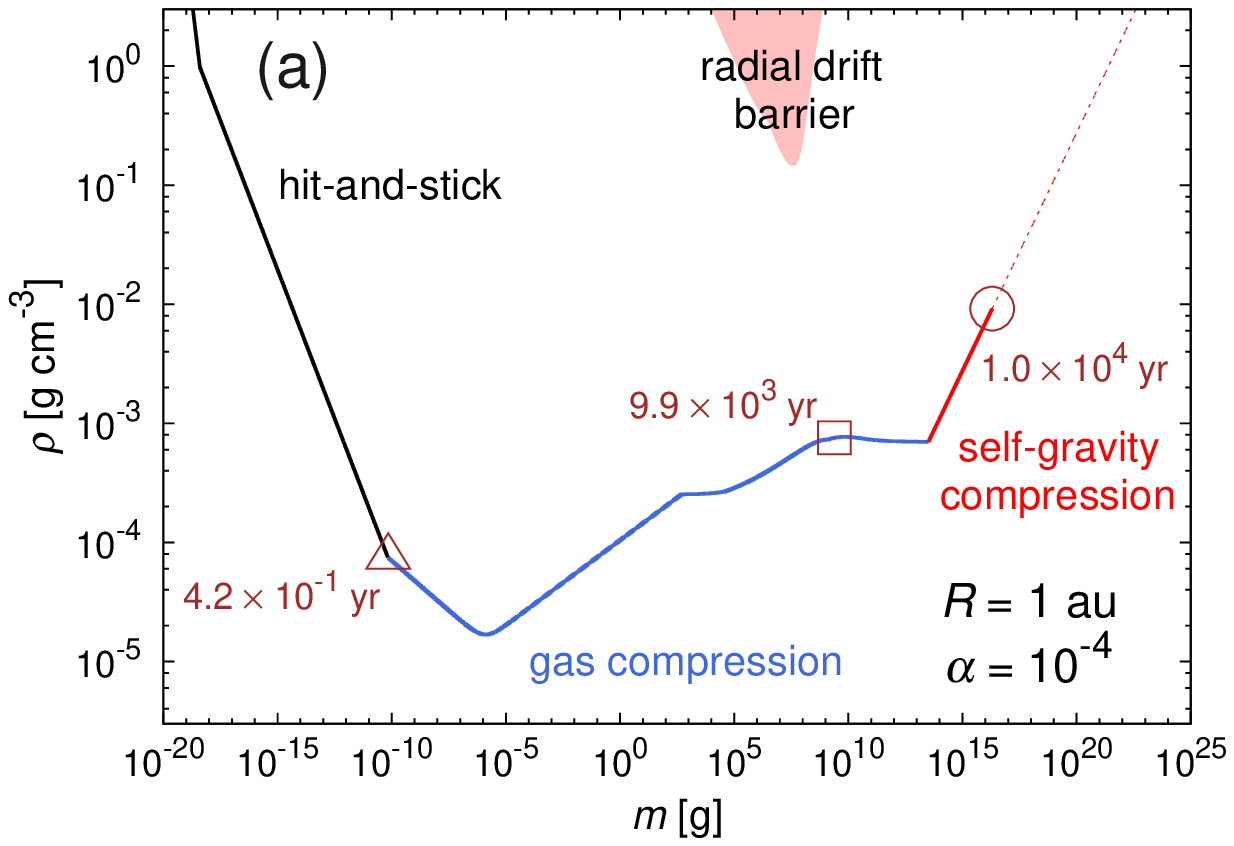}{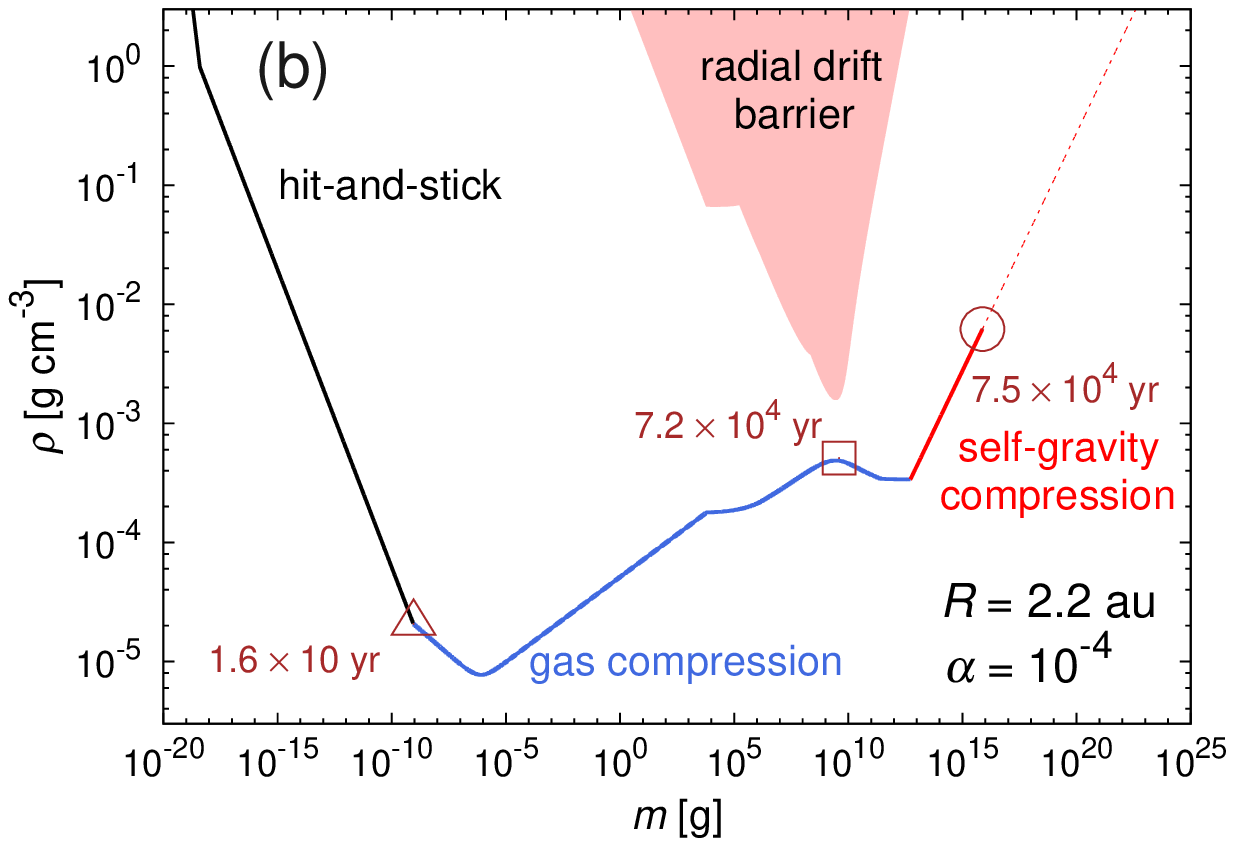}
\plottwo{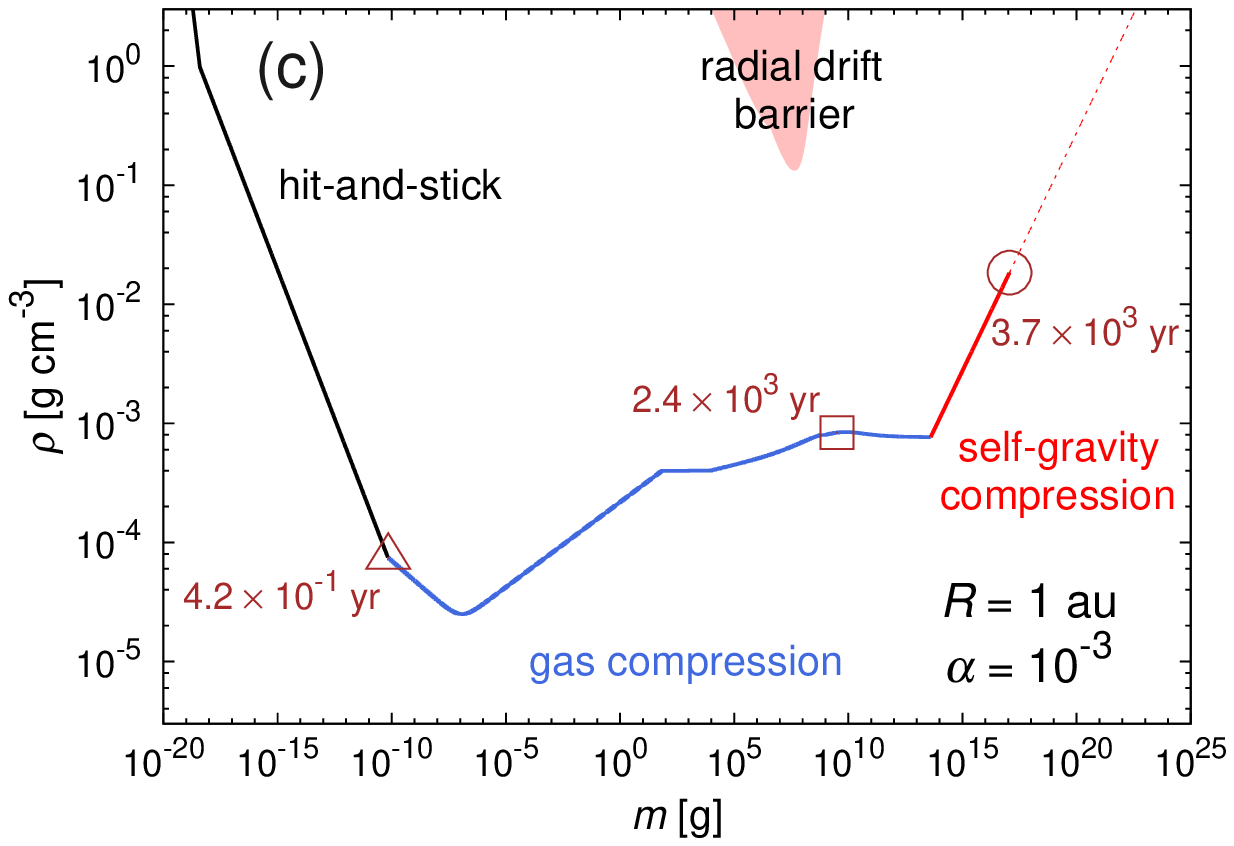}{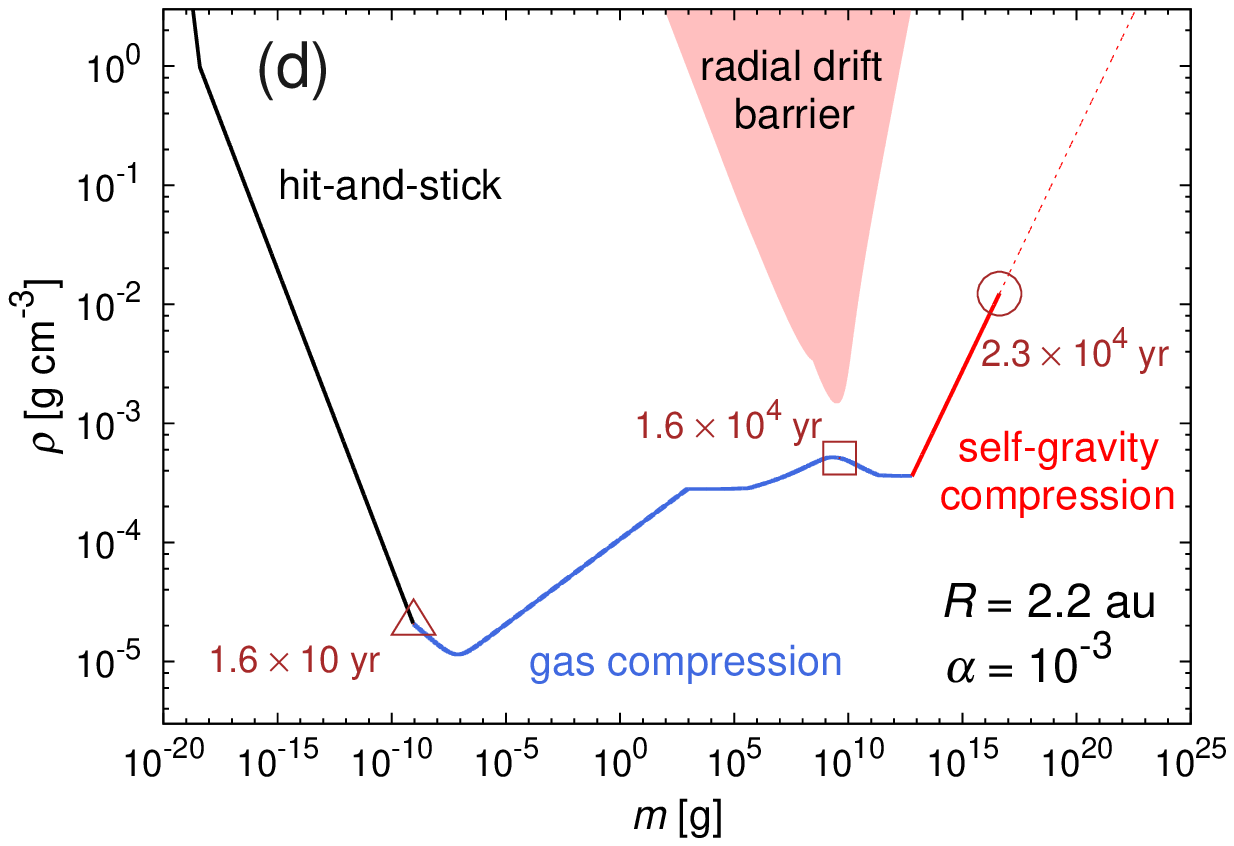}
\vspace{1.5truecm}
\caption{
Pathways of rocky planetesimal formation in the minimum mass solar nebula.
The black lines show the evolutional tracks through dust growth via hit-and-stick without compression.
The blue and red lines represent dust aggregation with gas compression and self-gravity compression.
The solid lines (black, blue, and red) show that dust aggregates evolve with orderly growth, and dashed lines (red) represent the runaway growth.
The triangles, squares, and circles mark the sizes for which $\rho_{\rm{hit}} = \rho_{\rm{gas}}$, $\Omega_{\rm{K}} t_{\rm{s}} = 1$, and ${\Delta v} = \sqrt{{G m } / r}$, respectively.
The pink shaded regions indicate where the timescale of radial drift is less than the timescale of growth.
(a): for $R = 1\ \rm{au}$ and $\alpha = 10^{-4}$.
(b): for $R = 2.2\ \rm{au}$ and $\alpha = 10^{-4}$.
(c): for $R = 1\ \rm{au}$ and $\alpha = 10^{-3}$.
(d): for $R = 2.2\ \rm{au}$ and $\alpha = 10^{-3}$.
}
\end{figure*}

For all the cases, our calculations show that the equilibrium density of statistic compression is higher than the density obtained by collisional compression.
Therefore, we initially obtained the pathways of dust growth from (i) fractal growth via hit-and-stick aggregation, then from (iii) gas compression, and finally from (iv) self-gravity compression.
We show that the revealed pathways overcome the radial drift problem whether the turbulence is weak ($\alpha = 10^{-4}$) or strong ($\alpha = 10^{-3}$).
We only show the results of the rocky planetesimal formation in the minimum mass solar nebula in Figure 1, however, it is also possible to form planetesimals in more massive disks.
Furthermore, even if the dust density of the disk is lower than that of the minimum mass solar nebula, rocky planetesimals might be formed around $1\ \rm{au}$.

We also plot the temporal evolution of the mass and the density of aggregates in Figure 1.
The triangles and circles represent the starting conditions of gas compression of dust aggregates, $\rho_{\rm{hit}} = \rho_{\rm{gas}}$, and the starting conditions of runaway growth, ${\Delta v} = \sqrt{{G m } / r}$, respectively.
Our calculations reveal that dust aggregates would initially compress within a year (for $1\ \rm{au}$) or a few decades (for $2.2\ \rm{au}$) and kilometer-sized rocky planetesimals might be formed within several tens of thousands of years after nanograin formation.
In addition, during planetesimal formation, dust aggregates exist for most of the time as aggregates in which $\Omega_{\rm{K}} t_{\rm{s}} \ll 1$ and $m \ll 10^{10}\ \rm{g}$.

Our results suggest that what we can observe in protoplanetary disks are not monomer grains but large and fluffy aggregates because the timescale of growth is extremely short when aggregates are smaller than $1\ {\mu{\rm{m}}}$.
Since fluffy dust aggregates are readily stirred up to the surface layer of protoplanetary disks, they may affect the visible/near-infrared scattered light image.
\citet{Mulders+2013} have shown that the faint and asymmetric brightness of total intensity of HD 100546 with the Hubble Space Telescope indicate that large dust grains exist at the surface layer.
In addition, \citet{Stolker+2016} showed that large dust grains in the surface layer might have an aggregate structure which prevents them from settling efficiently towards the midplane.
Although constraint on the monomer size from observations is still challenging, we could estimate the monomer size by considering more accurate light scattering formulae of fluffy dust aggregates proposed by \citet{Tazaki+2016}.
This radiative transfer model for fluffy dust aggregates suggests that a disk scattered light observation can be presumably explained by fluffy aggregates composed of nanometer-sized monomers (Tazaki et al., in prep., personal communication).

In the final stage of aggregation, the runaway growth of planetesimals begins when the escape velocity from very large dust aggregates exceeds the collisional velocity of these dust aggregates \citep[e.g.,][]{Wetherill+1989,Kobayashi+2016}.
Our calculations reveal that the runaway growth starts when the dust aggregates become as massive as the order of $10^{16}\ \rm{g}$.
This suggests that terrestrial (proto)planets might be formed from small ($r \lesssim 10\ \rm{km}$) planetesimals.
This result is consistent with the conclusion of \citet{Kobayashi+2013}, which insists that Mars was formed from planetesimals smaller than $10\ \rm{km}$ in radius, to explain its small mass and rapid formation timescale obtained from $^{182} \rm{Hf}$--$^{182} \rm{W}$ chronometry.

\section{Conclusions}
Mineralogical and cosmochemical evidence suggest that silicate grains in meteorites are not interstellar grains but condensates formed via evaporaion of dust in the early solar nebula.
In addition, the condensates might have originally been nanometer-sized grains, according to the size distribution of matrix grains in primitive chondrites.
Therefore, we propose a new scenario in which rocky planetesimals in our solar system were formed by aggregation of nanometer-sized grains, and these nanograins are produced via evaporation and recondensation of dust.

We showed that rocky planetesimals can be formed at $1\ \rm{au}$, the distance from the Sun to Earth, and $2.2\ \rm{au}$, the distance from the Sun to the Flora family asteroids (which are probably one of the origins of ordinary chondrites), via direct collisional growth of silicate nanograins.
In addition, this scenario will provide a suitable distribution of rocky planetesimals for (proto)planet formation.

Although there is some evidence suggesting that silicate grains in meteorites are not interstellar grains but condensates formed via evaporation processes, no one knows how these nanograins were formed in our solar nebula, and how the grain growth and solid-state recrystallization change size-frequency distributions of matrix grains from log-normal distributions to power-law distributions.
We will address these issues in future work.

\acknowledgments
We thank Kenji Homma, Satoshi Okuzumi, and Ryo Tazaki for fruitful discussions.
This work is supported by JSPS KAKENHI Grant Number 15K05266.









\end{document}